# Effect of Heterogeneous Mixing and Vaccination on the Dynamics of Anthelmintic Resistance: A Nested Model

Lorenzo Sabatelli*¤

Vaccine and Infectious Disease Institute, Fred Hutchinson Cancer Research Center, Seattle, Washington, United States of America

## Abstract

Anthelmintic resistance is a major threat to current measures for helminth control in humans and animals. The introduction of anthelmintic vaccines, as a complement to or replacement for drug treatments, has been advocated as a preventive measure. Here, a computer-based simulation, tracking the dynamics of hosts, parasites and parasite-genes, shows that, depending on the degree of host-population mixing, the frequency of totally recessive autosomes associated with anthelmintic resistance can follow either a fast dynamical regime with a low equilibrium point or a slow dynamical regime with a high equilibrium point. For fully dominant autosomes, only one regime is predicted. The effectiveness of anthelminthic vaccines against resistance is shown to be strongly influenced by the underlying dynamics of resistant autosomes. Vaccines targeting adult parasites, by decreasing helminth fecundity or lifespan, are predicted to be more effective than vaccines targeting parasite larvae, by decreasing host susceptibility to infection, in reducing the spread of resistance. These results may inform new strategies to prevent, monitor and control the spread of anthelmintic resistance, including the development of viable anthelmintic vaccines.

**Funding:** The author has no support or funding to report.

**Competing Interests:** The author has declared that no competing interests exist.

* E-mail: lsabatel@fhcrc.org

¤ Current address: Institute for Health Metrics and Evaluation, Department of Global Health, University of Washington, Seattle, Washington, United States of America

## Introduction

Under favorable ecological conditions, the use of anthelmintic drugs may result in the spread of drug resistant parasite strains that spontaneously occur or are imported into parasite populations [1,2,3]. Conventionally, alleles with frequency greater than 1% in a population are considered wild types, while those with frequency below 1% are considered mutants [4]. The spread of alleles is generally influenced by host-parasite regulatory mechanisms, such as density dependence [5], by the genetic make-up of resistant strains, and by the mechanism of transmission [6]. Genetics and transmission dynamics of the parasites play key roles in the emergence and spread of new strains (e.g. the transmission of packets of infectious larvae can explain the fast spread of fully recessive resistance to levamisole in the sheep nematode *Haemoncus contortus* [6]). The combined effects of these contributing factors remain largely unexplored.

The main control strategy against helminths is periodic chemotherapy using anthelmintics (e.g. albendazole, mebendazole, pyrantel pamoate and levamisole) [7,8,9,10,11]. Concerns about the variable efficacy of these drugs and the possible development of resistance [3,10,12] have stimulated efforts to develop new methods of control, including vaccines [13,14,15,16,17]. For instance, *Ancylostoma* secreted protein-2 of *Necator americanus* (*Na-ASP*-2) [16,18,19,20] has been shown to protect dogs and hamsters from infection when challenged with larvae of human hookworm. Other existing or candidate vaccines against helminths rely on different biological mechanisms. For example, Cathepsin L from the monoecious sheep parasite *Fasciola hepatica* reduces *F. hepatica* egg production and egg viability [21], while *Anti-H-gal-GP* interferes with the digestive process in *H. contortus* leading to premature death of the parasites [22]. None of these vaccines provides or is likely to provide full protection against helminth infection and morbidity (model I or "leaky" vaccine [23,24]). As a result, in most cases it will still be necessary to treat individuals for existing infections using anthelmintics. By reducing the frequency of drug treatment (and the resulting selection pressure) and the reproductive rate of the parasites, vaccines can potentially decrease the spread of alleles associated with drug resistance.

Here a computer-based simulation tracking the dynamics of hosts, parasites and parasite-genes is used to study the transition of a gene associated with drug-resistance from mutant (or imported) to wild type in a population of dioecious helminths, and the potential contribution of anthelmintic vaccines to containing the spread of autosomal genes associated with drug resistance. Central to this study are: i) the role of between-host mixing, and of the associated degree of parasite inbreeding, in shaping the dynamics of fully dominant and fully recessive anthelmintic resistance; ii) the impact of vaccines targeting different stages and biological functions of helminths and their influence on the rate of selection imposed by external chemotherapeutic interventions; and iii) the impact of parasite genetics and host-host transmission patterns on vaccine performance. Due to its epidemiological relevance human hookworm is used as a case-study here.

Hookworms (Ancylostoma duodenale and Necator americanus) are the most widespread human intestinal parasites. They are estimated to infect approximately 600 million individuals living in the tropics and sub-tropics. Adult worms live and mate in the small intestine of humans. Eggs exit the body in the feces and contaminate the soil, where larvae emerge and molt. The infective larvae are capable of penetrating the skin of a new host. The larvae migrate through the host tissue and blood vessels reaching lungs and trachea, from where they are swallowed before maturing into adults in the small intestine [11].

## Results

### Ecological and epidemiological scenario

The simulation reproduces the following scenario (Figure 1). Initially, only a subgroup of an endemically infected human population carries parasites with the gene of anthelmintic resistance. The introduction of resistant parasites in this subgroup can be due to human mobility (e.g. immigration) or infection from a common source. The subgroup is assumed to be the focus, from which drug resistant parasites may begin to spread to the general population. Because inbreeding may play an essential role in the spread of drug resistance (5), individual hosts in the focus are assumed to belong to the same infective contact network, nevertheless they do not need to live in close proximity, i.e to belong to the same household. Infective contacts, i.e. the transmission of parasites between hosts, typically occur when hosts spend time in a common place with poor or no sanitations and whose environmental conditions are suitable for egg hatching, larvae molt and host penetration. Ecological examples include: schoolchildren attending the same school, workers of a plantation or mine without appropriate sanitation facilities. Hosts in the focus are assumed to be exposed, on average, to the same risk of infection and to carry, on average, the same parasite burden as the general population in their community. They may mix homogeneously with the general population or mix preferentially with other hosts belonging to the focus (assortative mixing) [25,26]. For instance workers in a particular plantation might be infected by and infect preferentially their co-workers, but also family members (e.g. school-age children) and other members of their (hyperendemic) community. The average proportion ($\rho$) of parasites that plantation workers acquire from and transmitted to co-workers is inversely related to the degree of mixing of this subpopulation (the focus) with the rest of the community. No assumption is made on whether parasite transmission occurs in bundles or one at a time, but from the same source (e.g. the loamy soil area surrounding an infected household, a school or a plantation).

Two alternative interventions are considered and compared in their impact on drug resistance: annual treatment with anthelmintic drugs or, every five years, anthelmintic treatment followed by anthemintic vaccination. The epidemiological variable of interest is $FA$, the frequency of resistant alleles in the general population, i.e the population that does not belong to the initial focus of resistance. Here $FA$ is used to measures the spread of resistance, as a function of time and population mixing.

The key simulation parameters are reported in Table 1. The basic reproductive number $R_0$ is equal to 3, the adult worm mortality rate $\mu$ is equal to 0.2. The parameters of the (negative-binomial) distribution of worms at equilibrium are the mean worm burden $W=20$ and the clumping parameter $k=0.34$ [27]. This study assumes no acquired immunity to hookworm infection. Anthelmintic drugs are assumed to kill 80% [28] of susceptible worms and 1% of resistant worms and to be administered every 1 year or, if combined with vaccine, every 5 year. The vaccine efficacy ranges from 10% to 90%. The vaccine duration $T$ is equal to 5 years.

The face validity of the model was shown by simulating the epidemiological scenarios of published hookworm re-infection studies [29][30][31][32] and comparing model predictions with the observed dynamics. A substantial agreement was found between observed and simulated re-infection time-scales (see *Supplementary Materials and Methods S1*).

### The dynamics of resistant alleles

When resistance is associated with a rare recessive allele, there are two alternative dynamical regimes associated with the spread of resistant alleles outside the primary focus of resistance (Figure 2, A–B and Figure S1) depending on whether the proportion $\rho$ of transmission occurring from hosts to hosts within the focus, which is associated with parasite inbreeding, is below or above a "threshold value". For the parameter values in table 1, the threshold value is 69% (range: 66%–71%) (see also *Supplementary Materials and Methods S1*). If $\rho$ is less than 69%, low degree of inbreeding, the allele frequency in the parasite population grows fast and plateaus (to 0.27%–0.30% for the parameter values in table 1) within 2–10 years since start of treatment. The positive selection pressure exerted by the drug cannot compensate for the reduction in mating (due to mortality of drug susceptible parasites)

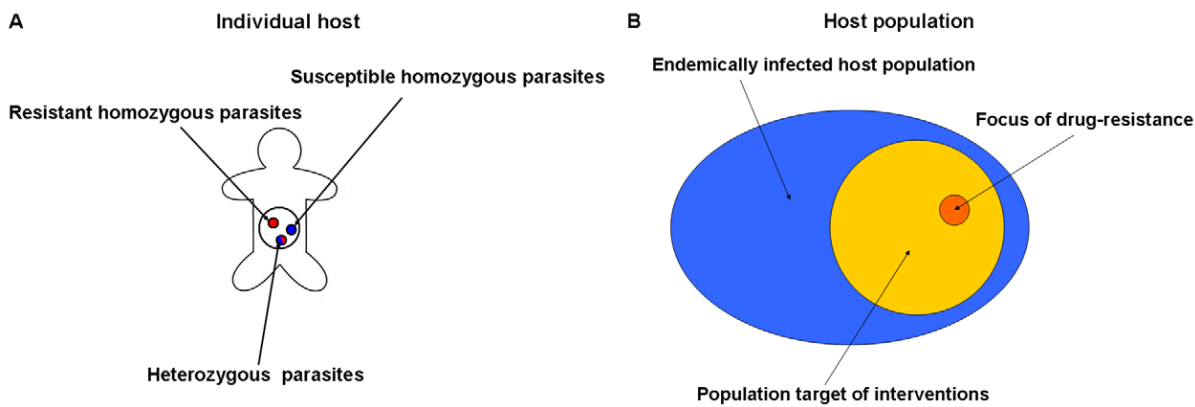

**Figure 1. Nested Host-Parasite-Gene populations.** A) A host may carry susceptible and resistant parasites, a parasite can be either homozygous or heterozygous for the gene associated with resistance; B) Euler diagram of the host population. Alleles associated with drug resistance are initially concentrated within a sub-set (focus) of the host sub-population (target) that receives either annual chemotherapy or chemotherapy and vaccine every $T$ (e.g. every 5) years.
doi:10.1371/journal.pone.0010686.g001

**Table 1.** Baseline parameters for the model.

| Parameter | Values | Description | Reference |
|---|---|---|---|
| $\alpha_{g,sus}$ | 80% | Anthelmintic efficacy in susceptible worm populations | [28] |
| $\alpha_{g,res}$ | 1% | Anthelmintic efficacy in resistant worm populations | Present study |
| $R_o$ | 3 | Basic reproductive number | Present study, based upon [1] |
| $\mu$ | 0.2/year | Worm mortality rate | [1] |
| W | 20 worms/person | Mean worm burden (intensity of infection) | Present study |
| k | 0.34 | Aggregation parameter (or clumping parameter) | [27] |
| $\Omega$ | 0 | Strength of naturally acquired immunity | Present study, based on: [1][11][27] |
| $\omega$ | <0.1/year | Immunological memory | Present study, based on: [1][11][27] |

doi:10.1371/journal.pone.0010686.t001

and for the 25% Mendelian constraint acting on the inheritance of the resistant phenotype. If $\rho$ is higher than 69%, indicating a high degree of inbreeding, a long initial phase (6–12 years) of slow growth is followed by a short phase of fast growth and then by a plateau, with the allele frequency reaching up to 1.7% 20 years after the start of treatment. If $\rho$ is about 69% a quasi-linear growth of the allele frequency is observed from year 2 to year 10. A non-monotonic relationship between the allele frequency at time $T = 20$ years and $\rho$ is observed, with a maximum allele frequency observed for $\rho = 85\%$.

The frequency of resistant alleles within the focus (see Figure S2): i) decreases monotonically with time and plateaus if $\rho<69\%$; ii) declines by about 25% within the first three years and grows thereafter if $\rho = 69\%$; or iii) grows quasi-monotonically if $\rho>69\%$.

To disentangle the effect of genetics and mixing from other aspects of the model, such as density dependence and population structure, and to understand the origin of the two-regime behavior of recessive resistant genes, a simple deterministic model of allele transmission and spread, based on a birth-death process [33], regulated by Mendel's laws [4] and subject to heterogeneous mixing, was developed (See *Supplementary Materials and Methods S1*). The deterministic model results suggest that (Figure S3), for sufficiently high levels of selective pressure and depending on the balance between production and dissemination of drug resistant alleles, the time dynamics of (recessive) allele frequency follows a two-regime behavior controlled by the parameter ρ, consistently with the results of the stochastic individual-based model used in this study.

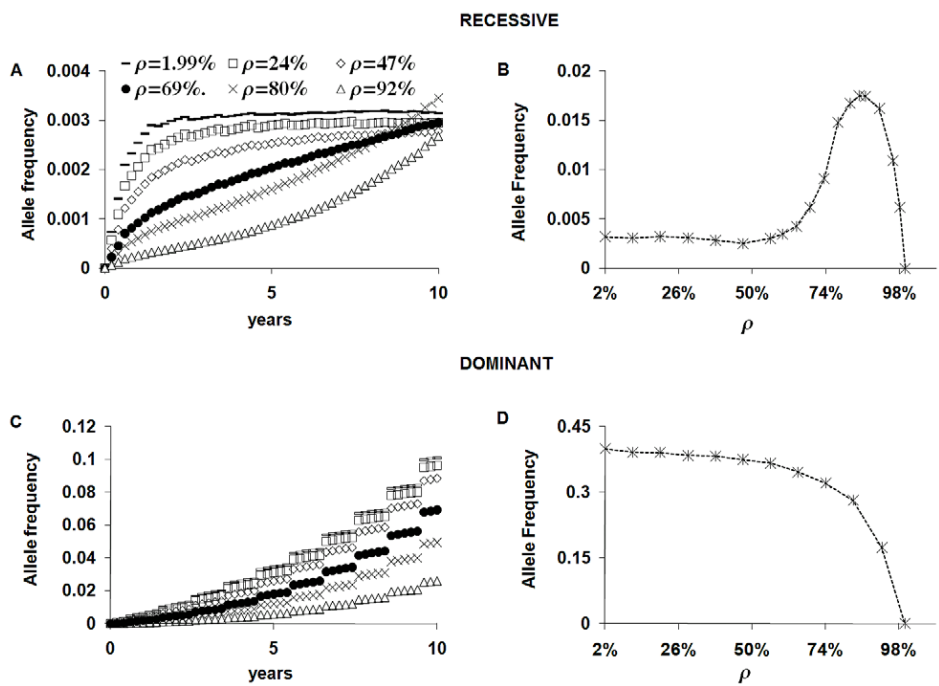

**Figure 2. Allele frequency in worms outside of the primary focus of resistance.** Resistance can be fully recessive (A–B) or fully dominant (C–D). Annual chemotherapy is administered to 50% of the host population and no vaccination. Simulation parameters in Table 1. A) If resistance is a recessive character the allele frequency follows two different time-dynamics depending on the the degree of host-population mixing $\rho$: i) for $\rho$ smaller than 69%, it grows fast and plateaus at a level inversely related to the value of $\rho$; ii) for $\rho$ equal to or larger than 69%, a slow initial growth is followed by an accelerated growth. B) Non-monotonic relation between the allele frequency after 20 annual chemotherapy rounds and $\rho$. The maximum allele frequency is observed when $\rho$ is equal to 85%. C) If resistance is a dominant character, the allele frequency grows at a rate inversely related to $\rho$. D) Negative monotonic relationship between the allele frequency after 20 annual chemotherapy rounds and $\rho$. The maximum allele frequency is observed when $\rho$ is equal to 1.99% (homogeneous mixing case).
doi:10.1371/journal.pone.0010686.g002

When resistance is associated with a fully dominant allele, invasion and strain replacement are steady processes, with a step-growth of the allele frequency (Figure 2, C–D). The degree of parasite inbreeding within the focus determines the rate at which the resistant strain will spread. Unlike in the case of recessive alleles, assortative mixing of resistant alleles slows the rate of invasion. For instance, if ρ = 92% the population invasion is almost four times slower than if the host-population mixes homogeneously, with respect to transmission, with the general population.

The dynamics of resistant alleles is essentially driven by their genetic dominance, by the degree of host-population mixing, and by the selective pressure exerted by drugs. The spread of resistance outside the initial focus is driven by two competing actions: production of offspring with the drug-resistant phenotype and dissemination of the drug-resistant offspring outside the initial focus. High transmission clustering increases the first, while reducing the second.

### The effect of vaccination

Vaccines can change the dynamics of resistant alleles (Figure S4) by stopping or slowing replacement of the susceptible strain with the resistant strain. Figure 3 presents the population-level vaccine effectiveness $VE$ (see eq. 6.1) as a function of the allele dominance and of the biological target of the vaccine. Here density-dependence acts on reproduction. $VE$ increases with gene dominance and is higher when drug-resistance is associated with a fully dominant gene. $VE$ is higher when the vaccine acts by increasing parasite mortality and by reducing parasite fecundity than when it acts by reducing host-susceptibility.

In the fully recessive case, $VE$ increases with ρ, in all situations. In the fully dominant case $VE$ increases with ρ, if the vaccine reduces host susceptibility and, slightly decreases with ρ, if the vaccine reduces female parasite fecundity; $VE$ is virtually independent of ρ, if the vaccine reduces the parasite lifespan. The results for density dependence acting on parasite establishment (Figure S5) are qualitatively similar in the recessive case, while they differ in the dominant case, for $VE$ increases with ρ, if the vaccine reduces the parasite lifespan.

The greatest net gain from combining vaccines with different modes of action (see for example Figure 4) is obtained when the single vaccine efficacies are low. In general, the effectiveness of two different vaccines (say *s* and *r*) are found to be sub-additive:

$$(1 - VE(s+r)) > (1 - VE(s))(1 - VE(r))$$

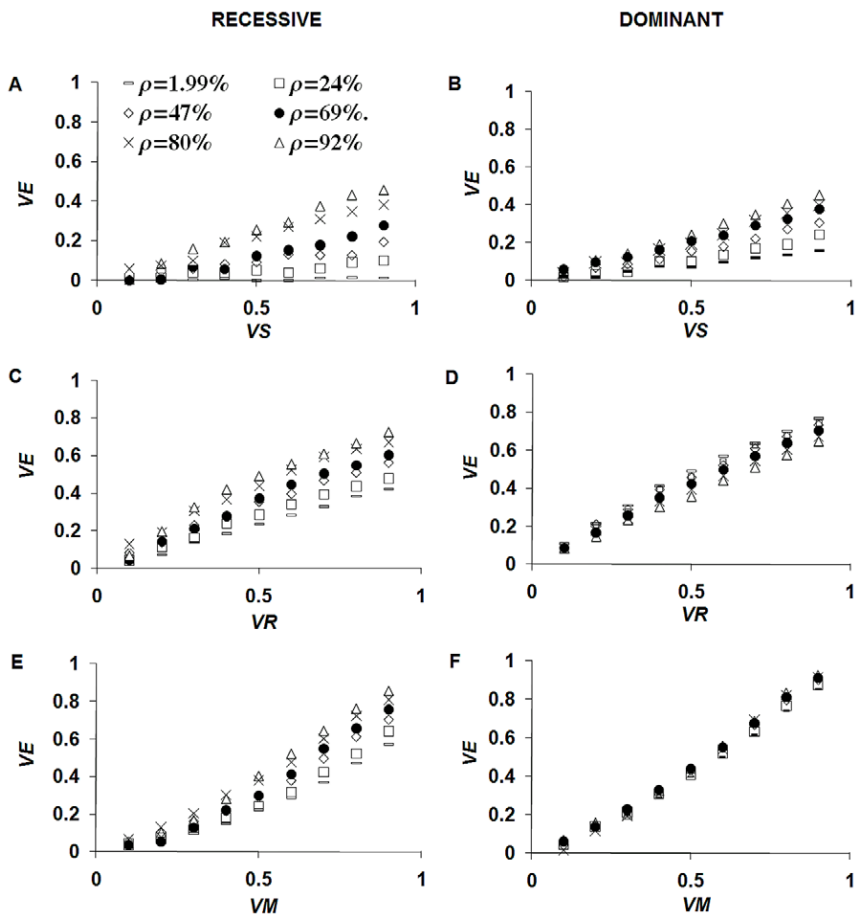

Figure 3. Population-level vaccine effectiveness (*VE*), assuming that density dependence acts on the fecundity of established parasites, and that 50% of the host population is effectively vaccinated (Simulation parameters in Table 1). Vaccines can reduce the host susceptibility to infection (A–B), by a proportion *VS*, or the female parasite reproductive rate (C–D), by a proportion *VR*, or the parasite lifespan (E–F), by a proportion *VM*. Vaccines reducing the parasite fecundity and vaccines reducing the parasite lifespan are more efficacious than vaccines reducing host susceptibility by the same proportion. Low reductions (<50%) in female reproductive rate have marginally higher impact than equal proportional reductions in adult worm lifespan. The impact on recessive alleles (A–C–D) grows with ρ, the degree of host-population mixing. The impact on dominant alleles increases with (B), decreases with (D) or is independent of (F) ρ, depending on the biological action of the vaccine.
doi:10.1371/journal.pone.0010686.g003

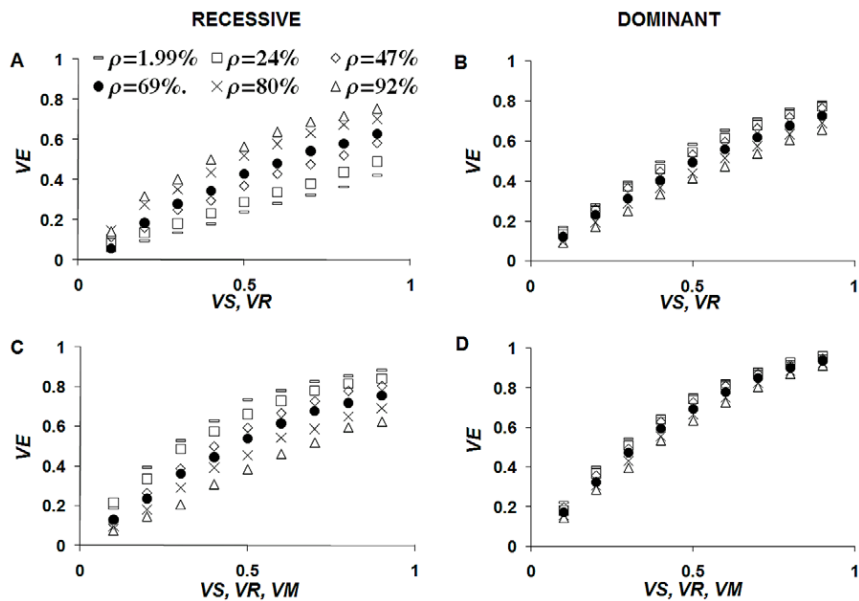

**Figure 4. Impact of multi-target vaccines.** Vaccines simultaneously reducing fecundity (*VR*) and susceptibility (*VS*) (panels A–B) or fecundity, susceptibility, and parasite life-span (*VM*) (panels C–D) have the greatest potential. In this example, it is assumed that the vaccine reduces by the same proportion female worm fecundity, host susceptibility, and parasite lifespan. A and C refer to recessive alleles, B and D refer to dominant alleles.
doi:10.1371/journal.pone.0010686.g004

### Sensitivity analysis

The sensitivity of the results to model parameters and assumptions was assessed through univariate and multi-variate analysis, as reported in Tables S1, S2, S3, S4. The spread of anthelmintic resistance is favored by high drug coverage, overdispersion of parasites, and a high basic reproductive number. The level of transmission clustering required for the transition from fast to slow dynamics in recessive populations decreases as drug coverage, overdispersion and the basic reproductive number increase. On the other hand, it increases if there is a residual drug susceptibility associated with the resistant phenotype.

The spread of resistant genes is most efficient in high prevalence settings, if the host life expectancy is significantly longer than the waning time of acquired immunity (if any) and if density-dependent constraints act on worm establishment rather than fecundity.

Reducing the frequency of drug treatment lowers the frequency of resistant recessive alleles only if $\rho > 69\%$, while it lowers the frequency of dominant alleles independently of the value of $\rho$.

The allele dynamics are sensitive to relative susceptibility of worm strains to anthelmintics, here measured as the ratio between the effectiveness of anthelmintics on the most susceptible and on the least susceptible strains. For recessive alleles, the results are more sensitive to the relative susceptibility when $\rho$ is above 69%.

The initial density of resistant worms in the focus has repercussions on the rate at which alleles spread in the population. Recessive alleles may go extinct and dominant alleles may temporarily decrease in number, if their frequency and numbers are too low, due to a low probability of worm mating and to phenotype inheritance constraints (due to Mendel's laws).

Higher vaccine coverage increases the protective effect if the vaccine reduces susceptibility to infection while it has relatively little impact if the vaccine reduces fecundity or increases worm mortality, provided that the focus of resistance is fully vaccinated. Increasing the basic reproductive number increases the impact of the vaccine relative to chemotherapy. This increase is small for highly efficacious vaccines, but rather large for low efficacious vaccines. Increasing parasite overdispersion also increases the impact of the vaccine relative to chemotherapy alone.

The impact of vaccines is marginally higher if the density dependence process acts on establishment rather than acting on fecundity (Figure S5).

### Discussion

Genetic dominance and patterns of contact between hosts are major determinants of the dynamics of anthelmintic resistance both in the presence and in the absence of vaccination.

Under the model assumptions, depending on selection pressure and on host-population mixing, the dynamics of recessive genes can be very fast and lead to equilibrium between resistant and susceptible parasite strains or may be very slow (and possibly go undetected for many years) before entering a phase of fast growth. Because the existence of fast and slow allele dynamics implies that different levels of mixing result in different time-scales for the diffusion of recessive allele, it can be argued that, in a real-world setting, characterized by a highly heterogeneous contact network structure, recessive resistance may spread through sudden jumps preceded and followed by long periods of equilibrium and coexistence with susceptible strains. Unlike the case of dominant genes, the spread of recessive genes is non-monotonically related to the degree of isolation of the focus from which resistant parasites spread to the general population. Validity and robustness of this result were independently tested using a genetically-structured deterministic model of helminth transmission (see *Supplementary*

*Materials and Methods S1*). A two-regime time-behavior and a non-monotonic relation between spread of recessive resistant genes and mixing were observed for a range of key biological parameters (e.g. basic reproductive number, natural parasite mortality rate, endemic mean worm burden and initial prevalence of resistant genes) and intervention-related parameters (e.g. coverage and drug-induced parasite mortality rate). Unlike in the main model, the frequency of resistant alleles does not usually plateau, as observed in the fast dynamical regime within the main model, but it rather grows for several decades at a very slow rate. This is due to the fact that the simple deterministic model does not track individual hosts and therefore does not account for reductions in the mating probability due to within-host segregation of same-sex parasites (i.e. some hosts will harbor same-sex parasites, not able to reproduce). In general, individual-based models, such as the one used in this study, are more suited to capture the dynamical features associated with within-host parasite segregation.

The dynamics of dominant alleles initially clustered within a small proportion of individual hosts are not qualitatively affected by the specific patterns of transmission.

Vaccines capable of reducing the adult parasite life span have the greatest potential to prevent the spread of resistant genes, followed by those reducing fecundity and susceptibility. However, vaccines that only reduce fecundity would appear rather unattractive to individuals, as they provide no direct benefit to those who are vaccinated. A multi-target vaccine, combining different single-target vaccines acting on different biological functions and life stages of the parasite, e.g. susceptibility to larva migration, reproduction and life span, would prove the most effective option, especially if none of the single-target vaccines are highly efficacious, and would prove more robust against antigenic drift [34]. These findings may elicit and contribute to scientific debates and operational decisions on the development and use of vaccines against helminthes and other macro-parasites.

The results presented in this study assume that resistant parasites spread to the general host population from a group of hosts representative of, but generally not mixing homogeneously with, the general host population. Under endemic equilibrium this implies a "zero-sum" transmission between the group of initial carriers (focus of resistance) and the general population. This assumption may not hold if resistant parasites are clustered in super-spreaders, i.e. individuals with medium to high parasite burden adopting behaviors that promote parasite spread (e.g. open field defecation), in which case recessive genes may spread faster than predicted in this study.

As with all models, this framework only provides a crude representation of hookworm transmission dynamics and could be expanded to capture additional features, such as acquired immunity and seasonality. However, to date, there is no clear understanding of and agreement on the role played by immunological processes in human hookworm infection [8,11,35,36]. Age-intensity curves suggest that, at a population level, the effect of acquired immunity is weak and short-lived [11]. Seasonal effects have not been included in the model because the relevant time scales of the model (frequencies of treatment and vaccination, and the parasite and host life-spans) are equal to or longer than one year, nevertheless seasonal phenomena, such as rain and drought, are likely to have an impact on the quantitative estimates provided here.

The model is built on our current understanding of helminth epidemiology [7,9,10,14,16,37,38,39] and its outputs are consistent with the re-infection time-scales observed in comparable field studies [29,30].

However, factors such as the number of genes involved in drug resistance, the way the genes interact, e.g. dominance, complementary action, heterogeneity, epistasis, and any potential fitness costs associated with resistance, e.g. due to gene pleiotropy [4], are likely to affect the quantitative predictions of the model.

The qualitative results presented here would hold with the addition of further biological and ecological complexity and may be sufficiently general to be valid for other dioecius macroparasitic pathogens, e.g. *Plasmodium falciparum*, characterized by density dependence, transmission heterogeneity and clustering [17,40,41], although, to date, there is no clear evidence of inbreeding for malaria [42].

Better, evidence-based, information on density-dependence, immunity, genetic proximity of in-host parasites and genetic determinants of predisposition to infection would allow the development of more detailed models and provide more robust estimates.

From a general standpoint, this work illustrates the role of host-to-host transmission heterogeneity on anthelmintic dynamics and suggests that vaccines may need to target more than one parasite life-stage or biological function to help control macro-parasitic diseases, in the absence of more radical and effective interventions (e.g. ensuring human access to clean water, sanitations, shoes, etc.) capable of altering the host-parasite ecology for the benefit of the hosts.

## Materials and Methods

A stochastic discrete-time and discrete-event model, combining meta-population-based and individual-based approaches, was developed to explore the dynamics of anthelmintic resistance and the impact of vaccination on resistant strains. The simulation tracks the gene inheritance within parasite meta-populations, defined by sex and genotype, harbored by individual hosts, which are characterized by age, susceptibility to helminth infection, host-to-host transmission patterns, and genotype- and sex-specific parasite burden.

### 1. Model Structure

The model simulates the distribution and dynamics of parasites harbored in individual hosts and of genes, associated with drug resistance, within the parasite population. Let $\underline{W}(t) = \{W_1(t)....W_N(t)\}$ denote the vector of individual worm burdens in a host (e.g. human) community at time $t$. Let $A_r(t) = \sum_{j=1}^{N} \sum_{s=1}^{2} \sum_{g=1}^{3} (g-1) W_{j,s,g}(t)$ and $FA_r(t) = A_r(t) \Big/ 2 \sum_{j=1}^{N} \sum_{s=1}^{2} \sum_{g=1}^{3} W_{j,s,g}(t)$ be the number of copies and the frequency of the allele associated with worm resistance within a worm population hosted by a host community of size $N$ at time t, with $s=1$ for female worms and $s=2$ for male worms. $W_{j,s,g}(t)$ is the sex $s$ and genotype $g$ specific worm burden of individual $j$, $W_j(t) = \sum_{g=1}^{3} W_{j,1,g}(t) + W_{j,2,g}(t)$ the total worm burden of individual j and $W(t) = \sum_{j=1}^{N} \sum_{g=1}^{3} \left( W_{j,1,g}(t) + W_{j,2,g}(t) \right)/N$ the mean worm burden.

The rate of acquisition of new worms in a host depends on the basic reproductive number of the worm $R_0$ (the number of female parasites produced by an average female parasite, when there are no density dependent constraints acting anywhere in the life-cycle of the parasite [1]), a community density-dependent constraint $f(.)$ [33,43,44], known as negative density-dependence (reducing female parasite fecundity as the number of parasites within a host increases the rate of acquisition of new parasites as the number of parasites within a host increases, consistently with empirical evidence [43]) and the susceptibility of host $j$ to the acquisition of new worms $\tilde{h}_j$ [39].

Then, $f(.)$ takes the form:

$$f_{j,g}^R(\gamma,\underline{W}) = \frac{1}{N}\sum_{i=1}^{N} Z_{j,i} C_{i,g} \frac{\sum_{g^l=1}^{3} W_{i,1,g^l}}{\left[\sum_{g'=1}^{3}\left(W_{i,1,g'}+W_{i,2,g'}\right)\right]^\gamma} \quad (1.1)$$

Where

$$Z_{j,i} = \begin{cases} \rho N/N_k & \text{if } i,j \in \text{Focus} \\ (1-\rho)N/(N-N_k) & \text{if } i \notin \text{Focus}; j \in \text{Focus} \\ (1-\rho)N/(N-N_k) & \text{if } i \in \text{Focus}; j \notin \text{Focus} \\ (N-(2-\rho)N_k)N/(N-N_k)^2 & \text{if } i \notin \text{Focus}; j \notin \text{Focus} \end{cases} \quad (1.2)$$

This choice of $Z_{j,i}$ preserves the negative binomial distribution of parasites across the host population while ensuring that a fraction $\rho$ of all the infections occurring within the focus is determined by the offspring of worms harbored by individual hosts in the focus.

$$C_{i,g} = \begin{cases} \frac{\sum_{g',g''=1}^{3} W_{i,1,g'} M_{g,g',g''} W_{i,2,g''}}{\left(\sum_{g^l=1}^{3} W_{i,1,g^l}\right)\left(\sum_{g^p=1}^{3} W_{i,2,g^p}\right)} & \text{if } \left(\sum_{g^l=1}^{3} W_{i,1,g^l}\right)\left(\sum_{g^p=1}^{3} W_{i,2,g^p}\right) > 0 \\ 0 & \text{if } \left(\sum_{g^l=1}^{3} W_{i,1,g^l}\right)\left(\sum_{g^p=1}^{3} W_{i,2,g^p}\right) = 0 \end{cases} \quad (1.3)$$

where

$$M_{1,g',g''} = \begin{pmatrix} 1 & 0.5 & 0 \\ 0.5 & 0.25 & 0 \\ 0 & 0 & 0 \end{pmatrix}; \quad M_{2,g',g''} = \begin{pmatrix} 0 & 0.5 & 1 \\ 0.5 & 0.5 & 0.5 \\ 1 & 0.5 & 0 \end{pmatrix};$$

$$M_{3,g',g''} = \begin{pmatrix} 0 & 0 & 0 \\ 0 & 0.25 & 0.5 \\ 0 & 0.5 & 1 \end{pmatrix} \quad (1.4)$$

are determined by the Mendelian inheritance laws [4] and express the likelihood of obtaining genotype 1, 2 or 3 by crossing genotypes $g'$ and $g''$.

$$\Lambda_j(t) = 1 - \Omega \sum_{d=t_{birth}}^{t} \sum_{g'=1}^{3} \left(W_{i,1,g'}(d) + W_{i,2,g'}(d)\right)/\exp[\omega(t-d)] \quad (1.5)$$

is the proportional reduction of susceptibility to infection due to acquired immunity [1]. With $\Omega$ representing the strength of acquired immunity and $\omega^{-1}$ the average immunological memory.

The adult worm mortality is $\mu$. Similarly, if density dependence acts upon the rate of establishment of adult worms

$$f_{j,g}^E(\varepsilon,\underline{W}) = \frac{e^{-\varepsilon\left(\sum_{s'=1}^{2}\sum_{g''=1}^{3} W_{j,s',g''}\right)}}{N}\sum_{i=1}^{N} Z_{j,i} C_{i,g}\left(\sum_{g'=1}^{3} W_{i,1,g'}\right) \quad (1.6)$$

where $\gamma$ is the power-law exponent and $\varepsilon$ is the rate of exponential decay.

Thus the average rate at which new worms of genotype $g$ and sex $s$ ($s=1$ or $s=2$) are acquired by host $j$ is given by: $\varphi_S \mu\ R_0 h_j \Lambda_j f_{j,g}$ ($\varphi_S$ is the fraction of worms of sex $s$, $\varphi_1+\varphi_2=1$) and are cleared in host $j$ at rate $\mu_j$, the parasite death rate.

If $\Delta W_{j,s,g}(t)$ denotes the genotype- and sex-specific change in worm burden in individual $j$ occurring between time $t$ and $t+\delta t$ ($\delta t$ being the time-step of the simulation) then an individual's worm burden dynamics are described by the equations:

$$\begin{aligned}\Delta W_{j,s,g}(t) &= \tilde{p}_{j,s,g}(t) - \tilde{d}_{j,s,g}(t) \\ \tilde{p}_{j,s,g}(t) &\sim Poisson\left(\varphi_S \mu\ R_0 h_j \Lambda_j I_j f_{j,g}\delta t\right) \\ \tilde{d}_{j,s,g}(t) &\sim Binomial\left(W_{j,s,g}(t), 1-e^{-\mu\delta t}\right)\end{aligned} \quad (1.7)$$

## 2. Endemic distribution of parasites

When hookworm is endemic in a community the endemic distribution of individual worm burdens $W_j = \sum_{g=1}^{3}\left(W_{j,1,g}+W_{j,2,g}\right)$ is best described by a negative binomial distribution whereby relatively few people harbor most of the parasites. This is built into the model by assuming that hosts have different levels of susceptibility $\tilde{h}_j$ ($\tilde{h}_j=h_j\Lambda_j$ and $h_j=\tilde{h}_j/\Lambda_j(0)$). Drawing the sex- and genotype-specific establishment rate $\Delta W_{j,s,g}(t)$ for a host $j$ from a Poisson distribution with a mean equal to $\tilde{h}_j\Delta W_{j,s,g}(t)$, the susceptibility $\tilde{h}_j$ from a Gamma distribution with variance $1/k$ and describing the dynamics of the mean parasite burden $W$ with an immigration-death process (with a birth rate $\frac{1}{2}\mu\ R_0 h_j \Lambda_j f_{j,g}^R(\gamma,\underline{W})$ and a death rate $\mu$), ensures that, at equilibrium, the individual distribution of parasites can be described by a negative binomial with mean $W$ and variance $\sigma^2 = W + \frac{W^2}{k}$ [8,43,45]. The initial value of $\Lambda_j(0)$ is calculated as a function of the age and worm burden $W_j$ of individual $j$. In the absence of drug treatment and ignoring possible seasonal variations (since the lifespan of a hookworm typically exceeds one year) and when the infection is at endemic equilibrium, the total worm burden in the community is constant. The parameter $\gamma$ and $\varepsilon$ are entirely determined by $R_0$, $k$, $W$ and by the choice of the appropriate functional relationship (either (1.1) or (1.6)). They can be derived by imposing the stationarity condition $\sum_{i=1}^{N} \Delta W_i(t=0) = 0$ [8]. Because the pre-intervention worm burden is assumed to be the same inside and outside the focus, the stationarity condition implies that the transmission from and to the focus is a stochastic process with expectation value equal to zero (zero sum game).

## 3. Demographic structure and age-intensity relationship

Individuals in the model live from 0 to 50 years and die thereafter. The mean worm burden increases as hosts become older and plateaus after the age of 20 [11].

## 4. Chemotherapy

If chemotherapy is introduced at time $t_c$ the worm burden of treated individuals will change as:

$$W_{j,s,g}(t_c) \rightarrow W_{j,s,g}(t_c) - \tilde{b}_{j,s,g} \quad (4.1)$$

$$\tilde{b}_{j,s,g} \sim Binomial\left(W_{j,s,g}(t_c), \alpha_g\right) \quad (4.2)$$

where $\tilde{b}_{j,s,g}$ is the number of worms killed by chemotherapy within an individual harboring $W_{j,s,g}(t_c)$ worms at time $t_c$ and $\alpha_g$ is the probability that an adult worm of genotype $g$ (which can be susceptible or resistant) is killed by chemotherapy.

## 5. Vaccination

Three different types of vaccine are considered:

i) A vaccine reducing the individual susceptibility to infection in host $j$ with susceptibility $h_j$ at time $t$ by a proportion $VS_j(t)$.

Following vaccination, susceptibility to infection in host $j$ becomes:

$$h_j \rightarrow h_j(1 - VS_j(t)) \quad (5.1)$$

ii) A vaccine reducing the female worm reproductive rate, in host $j$ with susceptibility $h_j$ at time $t$, by a proportion $VR_j(t)$.

$$C_{j,g} \rightarrow C_{j,g} = \begin{cases} (1-VR_j(t)) \frac{\sum_{g',g''=1}^{3} W_{i,1,g'} M_{g,g',g''} W_{i,2,g''}}{\left(\sum_{g^l=1}^{3} W_{i,1,g^l}\right)\left(\sum_{g^p=1}^{3} W_{i,2,g^p}\right)} & \text{if } \left(\sum_{g^l=1}^{3} W_{i,1,g^l}\right)\left(\sum_{g^p=1}^{3} W_{i,2,g^p}\right) > 0 \\ 0 & \text{if } \left(\sum_{g^l=1}^{3} W_{i,1,g^l}\right)\left(\sum_{g^p=1}^{3} W_{i,2,g^p}\right) = 0 \end{cases} \quad (5.2)$$

iii) A vaccine reducing the mean worm lifespan by a proportion $VM_j(t)$, and therefore increasing the parasite mortality rate:

$$\mu_j \rightarrow \mu_j/(1 - VM_j(t)) \quad (5.3)$$

The vaccine efficacy (called $V$ if does not refer only to a specific type of vaccine) wanes exponentially so that, on average, $V(t) = V(0)e^{-t/T}$, and where $T$ is the average duration of protection and, $V(0)$ is the direct proportional reduction in susceptibility to infection ($VS(0)$), female worm reproductive rates ($VR(0)$) or worm lifespan ($VM(0)$) that would be measured in a population in the absence of density dependent effects or waning of vaccine efficacy. Waning is implemented in the model by allowing "individual vaccine protection" to fail (i.e. to change from $VS(0)$ or $VR(0)$ to zero) at any one time step $\delta t$ with probability $\delta t/T$. Alternative mechanisms, such as a gradual reduction of individual protection (e.g. $VS_j(h,t) = VS_j(h,0)e^{-t/T}$) were also explored but did not substantially affect the results.

## 6. Vaccine Effectiveness

The vaccine effectiveness $VE$ is defined as the vaccine-induced proportional reduction of the frequency of the allele associated with drug resistance, within the human population of interest, over a period $T_f$

$$VE = 1 - \frac{\sum_{t=t_{1V}}^{T_f} FA_r(t,V)}{\sum_{t=t_{1V}}^{T_f} FA_r(t,U)} \quad (6.1)$$

where $FA_r(t,V)$ and $FA_r(t,U)$ are the frequencies of the allele associated with resistance, in the presence and in the absence of vaccination ($V$ stands for vaccinated and $U$ stands for unvaccinated). The first vaccination round starts at time $t_{1V} = 0$.

## 7. Simulation

The model was used to simulate the effect of chemotherapy alone and chemotherapy in combination with a vaccine allocated to 50% of the whole population.

The initial frequency of resistant alleles is equal to 1% (a borderline scenario between "mutant" and "wild") and is clustered in 1.99% of the human population (equal to the $2pq+q^2$ term from the Hardy-Weinberg equation $p^2+2pq+q^2$), to increase the mating probability of worms carrying the resistant gene. The proportion of transmission happening within this initial cluster (focus) of resistance $\rho$ ranges from 1.99% (random mixing) to 92% (highly heterogeneous mixing). The simulation reproduces the effects of annual chemotherapy treatments as well as the effects of combined chemotherapy and vaccination, administered every $T$ years, on susceptible and resistant worms and on the underlying allele populations, over a 10-year period.

The mathematical relationship between host age and mean worm burden [11] is consistent with a very weak and rapidly waning naturally acquired immunity, with strength $\Omega < 0.003$ and immunological memory $1/\omega < 3$ years. For the purposes of the baseline simulation, and without loss of generality, no acquired immunity to hookworm infection was assumed, nevertheless the impact of the naturally acquired immunity on the outcomes of this study was assessed as a part of the sensitivity analysis.

For each scenario, a mean statistic over 1,000 realizations, each of which involves 10,000 individuals followed over a period of 10 (or 20) years was computed. The demographic structure of the population and the parasite epidemiology are assumed to remain unchanged during the simulation. The time-evolution was attained via synchronous updating of the worm burden of each individual and of the force of infection at each run, assuming that contemporaneous infective episodes involving different individuals are independent. The time-step between two runs of the simulation is $\delta t = 0.2$ years. This is a trade-off between a high resolution time-dynamics, requiring a time-step much smaller than the average life span of the parasite ($\delta t \ll 1/\mu_0$), and the constraints set by the Markovian structure of the model, which overlooks the intermediate stages of the worm life-cycle (for hookworm, it takes approximately 9–12 weeks for an egg deposited in the soil to mature into a sexually mature adult [46]).

## Supporting Information

**Materials and Methods S1**
Found at: doi:10.1371/journal.pone.0010686.s001 (0.11 MB DOC)

**Figure S1** Frequency of resistant alleles after 10 years of chemotherapy. A) Reducing the frequency of treatment does not reduce the frequency of recessive alleles, unless the focus of resistance is highly isolated ($\rho = 92\%$). B) Reducing the frequency of treatment always reduces the frequency of dominant alleles. The frequencies of (C) recessive and (D) dominant alleles increase with the relative susceptibility to drugs, defined as the ratio of the probabilities of being killed by drugs for susceptible and resistant worms. Baseline parameters are in Table 1 of the main paper.
Found at: doi:10.1371/journal.pone.0010686.s002 (4.92 MB TIF)

**Figure S2** Frequency of recessive alleles within the focus.

Found at: doi:10.1371/journal.pone.0010686.s003 (4.03 MB TIF)

**Figure S3** Results from a simple deterministic model. A) If the anthelmintic associated rate of death is higher than 0.35 the allele frequency follows two different time-dynamics depending on the degree of host-mixing ρ: i) it grows fast and then very slow ii) a slow initial growth is followed by an accelerated growth. B) Non-monotonic relation between the allele frequency after 20 annual chemotherapy rounds and ρ obtained for different values of δ. The difference in allele frequency between the two dynamical regimes decreases as the selection-pressure-related parameter δ increases. Baseline parameters are in Table 1 of the main paper.
Found at: doi:10.1371/journal.pone.0010686.s004 (2.61 MB TIF)

**Figure S4** Effect of vaccination on the dynamics of drug resistance. In this example the vaccine reduces by a proportion VR the average reproductive rate of female parasites. A) Recessive case: the reduction in the spread of recessive resistant alleles primarily depends on the efficacy of the vaccine (in this example, the vaccine acts on female worm fecundity and ρ = 85%). B) Dominant case: the reduction in the spread of dominant recessive alleles primarily depends on the reduction of the frequency of treatment (in this example, every 5 years instead of every year). Baseline parameters are in Table 1 of the main paper.
Found at: doi:10.1371/journal.pone.0010686.s005 (2.23 MB TIF)

**Figure S5** Vaccine effectiveness (VE) when density dependence acts on the establishment of new worms. Vaccines can reduce the host susceptibility to infection (A–B), by a proportion VS, or the female parasite fecundity (C–D), by a proportion VR, or the parasite lifespan (E–F), or by a proportion VM. Vaccines reducing the parasite fecundity and vaccines reducing the parasite lifespan are more efficacious than vaccines reducing host susceptibility by the same proportion.
Found at: doi:10.1371/journal.pone.0010686.s006 (6.01 MB TIF)

**Table S1** Univariate sensitivity analysis: Threshold for transition ρ from fast to slow dynamics of recessive alleles. The value of ρ, estimated at baseline, is 69%. Intervals refer to maximum and minimum ρ values. Impact of a vaccine reducing host susceptibility for VS = 50% and ρ. Intervals of maximum and minimum percentage variation of vaccine impact VE, from baseline values VE = 18% (recessive) and VE = 17% (dominant) observed when parameter values vary within the given ranges. Baseline parameters are in Table 1 of the main paper.
Found at: doi:10.1371/journal.pone.0010686.s007 (0.04 MB DOC)

**Table S2** Multivariate Sensitivity Analysis: Variation of the threshold for transition ρ from fast to slow dynamics of recessive alleles. The value of ρ, estimated at baseline, is 69%. The density-dependent parameters are chosen to fit the endemic equilibrium data. Baseline parameters are in Table 1 of the main paper.
Found at: doi:10.1371/journal.pone.0010686.s008 (0.04 MB DOC)

**Table S3** Sensitivity analysis of the impact of a vaccine reducing host susceptibility for VS = 50% and ρ = 69%. Intervals of percentage variation of vaccine impact VE, from baseline values VE = 18% (recessive) and VE = 17% (dominant) observed when parameter values vary within the given ranges. The density-dependence parameters are chosen to fit the endemic mean worm burden W Min and Max are the minimal and maximal percentage deviation from the simulation results obtained using baseline parameters, when density-dependent regulatory mechanisms act on parasite fecundity. Baseline parameters are in Table 1 of the main paper.
Found at: doi:10.1371/journal.pone.0010686.s009 (0.04 MB DOC)

**Table S4** Sensitivity Analysis of the impact of a vaccine reducing female parasite fecundity, for VR = 50% and ρ = 69%. Intervals of percentage variation of vaccine impact VE, from baseline values VE = 18% (recessive) and VE = 17% (dominant) observed when parameter values vary within the given ranges. The density-dependence parameters are chosen to fit the endemic mean worm burden W. Min and Max are the minimal and maximal percentage deviation from the simulation results obtained using baseline parameters, when density-dependent regulatory mechanisms act on parasite fecundity. Baseline parameters are in Table 1 of the main paper.
Found at: doi:10.1371/journal.pone.0010686.s010 (0.05 MB DOC)

### Acknowledgments

The author wishes to thank M. Elizabeth Halloran, Dobromir T. Dimitrov, Nicole Basta and an anonymous reviewer for helpful comments.

### Author Contributions

Conceived and designed the experiments: LS. Performed the experiments: LS. Analyzed the data: LS. Contributed reagents/materials/analysis tools: LS. Wrote the paper: LS.